# Super-ASTROD: Probing primordial gravitational waves and mapping the outer solar system

**Wei-Tou Ni**[1,2]

[1]Center for Gravitation and Cosmology, Purple Mountain Observatory,
Chinese Academy of Sciences, Nanjing, 210008 China
[2]National Astronomical Observatories, Chinese Academy of Sciences, Beijing,
100012 China

E-mail: wtni@pmo.ac.cn

**Abstract**. Super-ASTROD (Super Astrodynamical Space Test of Relativity using Optical Devices, or ASTROD III) is a mission concept with 4 spacecraft in 5 AU orbits together with an Earth-Sun L1/L2 spacecraft ranging optically with one another to probe primordial gravitational waves with frequencies 0.1 μHz - 1 mHz, to test fundamental laws of spacetime and to map the outer-solar-system mass distribution and dynamics. In this paper we address its scientific goals, orbit and payload selection, and sensitivity to gravitational waves.

## 1. Introduction

The general concept of ASTROD (Astrodynamical Space Test of Relativity using Optical Devices) is to have a constellation of drag-free spacecraft navigate through the solar system and range with one another using optical devices to map the solar-system gravitational field, to measure related solar-system parameters, to test relativistic gravity, to observe solar g-mode oscillations, and to detect gravitational waves (GWs). Distances between drag-free spacecraft depend critically on solar-system gravity (including gravity induced by solar oscillations), underlying gravitational theory and incoming gravitational waves [1, 2]. A precise measurement of these distances as a function of time will enable the causes of temporal variation to be determined. The astrodynamical equation for spacecraft and solar-system bodies is

$$\ddot{\mathbf{r}}_i = -\sum_{j \neq i} \mu_j (\mathbf{r}_{ij}/r_{ij}^3) + \Delta\ddot{\mathbf{r}}_i(\text{1PN}) + \Delta\ddot{\mathbf{r}}_i(\text{2PN}) + \Delta\ddot{\mathbf{r}}_i(\text{GW}) + \text{gal-cosmo term} + \text{non-grav term}$$

$$(i = 0, 1, ..., n) \quad (1)$$

where the first term on right of (1) is Newtonian gravitational acceleration, the second term $\Delta\ddot{\mathbf{r}}_i(\text{1PN})$ [3] is

$$\Delta\ddot{\mathbf{r}}_i(\text{1PN}) = (1/c^2) \sum_{j \neq i} \mu_j (A_{ij}\mathbf{r}_{ij} + B_{ij}\dot{\mathbf{r}}_{ij}) \quad (2)$$

with

$$A_{ij} = \dot{r}_i^2/r_{ij}^3 - (1+\gamma)\dot{r}_{ij}^2/r_{ij}^3 + 3(\mathbf{r}_{ij}\cdot\dot{\mathbf{r}}_j)^2/2r_{ij}^5 + [2(\gamma+\beta)(\mu_i+\mu_j)+\mu_i]/r_{ij}^4$$
$$+ \sum_{k\neq i,j} \mu_k[2(\gamma+\beta)/r_{ij}^3 r_{ik} + (2\beta-1)/r_{ij}^3 r_{jk} + 2(\gamma+1)/r_{jk}^3 r_{ij}$$
$$- (2\gamma+1.5)/r_{jk}^3 r_{ik} - (\mathbf{r}_{ij}\cdot\mathbf{r}_{ik})/2r_{jk}^3 r_{ij}^3] \quad (3)$$
$$B_{ij} = \mathbf{r}_{ij} \cdot [2(\gamma+1)\dot{\mathbf{r}}_{ij} + \dot{\mathbf{r}}_j]/r_{ij}^3$$

and $\mathbf{r}_i = \mathrm{x}_i$, $\mathbf{r}_{ij} = \mathrm{x}_i - \mathrm{x}_j$, $m_i = GM_i/c^2$, $M_i$'s the masses of the *i-th* bodies, the third term is the 2PN term under current study, the fourth term is the acceleration due to GWs, the fifth term is the acceleration due to our galaxy and cosmos, and the last term is the acceleration due to non-gravitational origin. Integration of (1) gives the range as a function of time. When the spacecraft is drag-free, the last term is absent and all other terms can be determined from range measurement of geodetic motions; in the determination, spacecraft initial-position and initial-velocity parameters, solar-system parameters, relativistic-gravity parameters are estimated and their uncertainties determined (just like in the LLR [Lunar Laser Ranging] practice), while the residuals are analyzed for the (weaker) GW signals.

A baseline implementation (ASTROD II or simply ASTROD) of the general ASTROD mission concept is to have two spacecraft in separate solar orbits, each carrying a payload of a proof mass, two telescopes, two 1 - 2 W lasers, a clock and a drag-free system, together with a similar spacecraft near Earth at one of the Lagrange points L1/L2. The three spacecraft range coherently with one another using lasers to map solar-system gravity, to test relativistic gravity to 1 part per billion, to observe solar g-mode oscillations, and to detect gravitational waves in the frequency range 1 μHz-10 mHz [1].

A simplified implementation (ASTROD I) of one spacecraft optical ranging with Earth ground stations will be able to achieve a significant part of the goals in testing relativity and mapping the inner solar system, and yet provide an incremental technology demonstration [2].

For detection of primordial gravitational waves in space, one may go to frequencies lower or higher than the LISA [4] / ASTROD [1] bandwidth where there are potentially less foreground astrophysical sources to mask detection. DECIGO [5] and Big Bang Observer [6] look for gravitational waves in the higher range while Super-ASTROD looks for gravitational waves in the lower range. In the third phase, Super-ASTROD (ASTROD III), 4 spacecraft with 5 AU orbits together with an Earth-Sun L1/L2 spacecraft can be implemented to probe primordial gravitational waves with frequencies 0.1 μHz - 1 mHz and to map the outer-solar-system mass distribution and dynamics.

As to the test of relativistic gravity, we summarize the scientific goals of ASTROD and ASTROD I in Table 1. The first column lists the effect or quantity to be measured. The second column lists the present accuracy of measurement [7]. The third and fourth columns list the aimed accuracies of ASTROD I [2] and ASTROD [1] respectively. With larger orbits, the main goal of Super-ASTROD in test relativistic gravity is not to improve on PPN (Parametrized Post-Newtonian) parameters over ASTROD I / ASTROD, instead it is to test cosmological theories which give larger modifications from general relativity for larger orbits.

The GW sensitivity to primordial stochastic background is shown in Fig. 1 together with other gravitational detectors (see section 3 for explanation). The minimum detectable intensity of a stochastic GW background is proportional to detector noise spectral power density time frequency to the third power [4, 8]. Hence, with the same strain sensitivity, lower frequency detectors have an $f^{-3}$-advantage over the higher frequency detectors. This is the main reason why ASTROD and Super-ASTROD can probe deep into inflation GW region.

**Table 1** Summary of the scientific objectives in testing relativistic gravity of the ASTROD I and ASTROD missions

| Effect/Quantity | Present accuracy | Aimed accuracy of ASTROD I | Aimed accuracy of ASTROD |
|---|---|---|---|
| PPN parameter $\beta$ | $2 \times 10^{-4}$ | $3 \times 10^{-8}$ | $1.2 \times 10^{-9}$ |
| PPN parameter $\gamma$ (Eddington parameter) | $4.4 \times 10^{-5}$ | $3 \times 10^{-8}$ | $1 \times 10^{-9}$ |
| $(dG/dt)/G$ | $10^{-12}$ yr$^{-1}$ | $3 \times 10^{-14}$ yr$^{-1}$ | $5 \times 10^{-15}$ yr$^{-1}$ |
| Anomalous Pioneer acceleration $A_a$ | $(8.74 \pm 1.33) \times 10^{-10}$ m/s$^2$ | $0.7 \times 10^{-16}$ m/s$^2$ | $0.2 \times 10^{-16}$ m/s$^2$ |
| Determination of solar quadrupole moment parameter $J_2$ | $1 - 3 \times 10^{-7}$ | $1 \times 10^{-9}$ | $5 \times 10^{-11}$ |
| Measuring solar angular momentum via solar Lense-Thirring Effect | (0.1 for Earth) | 0.1 | better than $10^{-3}$ |
| Determination of planetary masses and orbit parameters | (depends on object) | 1 - 3 orders better | 2 - 4 orders better |
| Determination of asteroid masses and density | (depends on object) | 2 - 3 orders better | 3 - 4 orders better |

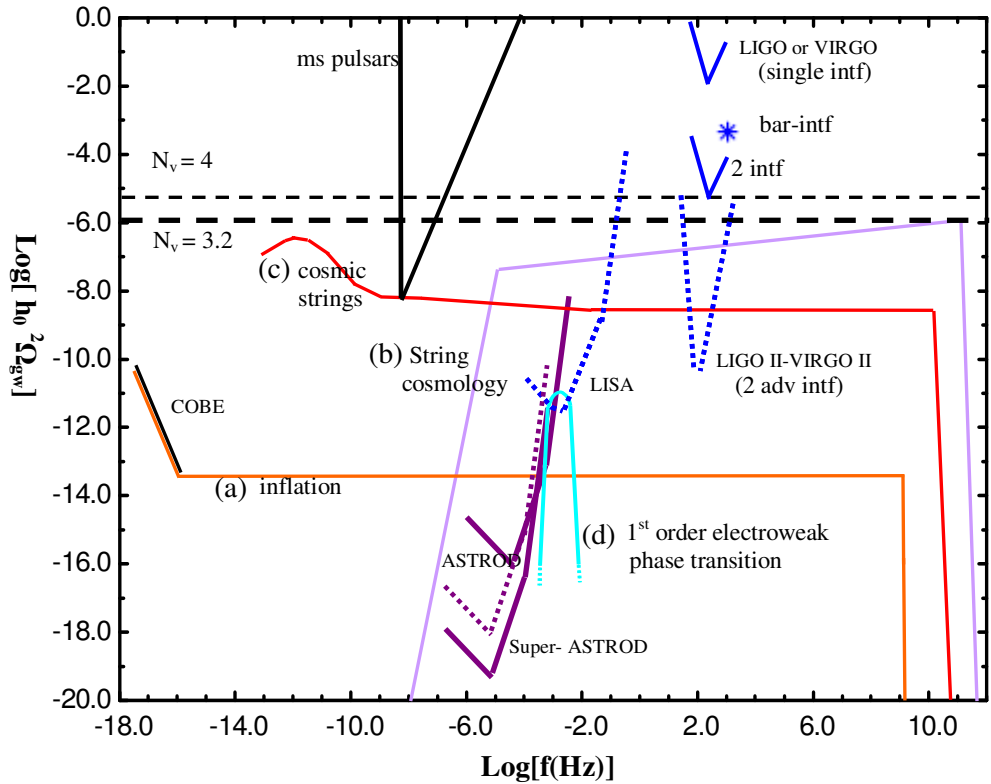

**Figure 1.** The stochastic backgrounds optimistically predicted with bounds and GW detector sensitivities. (Adapted from Figure 3 and Figure 4 of [8] with sensitivity curves of ASTROD and Super-ASTROD added; see [8] and section 3 for explanations).

In the following sections, we address to orbit selection, payload selection, sensitivity to gravitational waves, and scientific goals.

## 2. Mission concept

In 1996, we proposed Super-ASTROD in the first TAMA Meeting [9]: "With the advance of laser technology and the development of space interferometry, one can envisage a 15 W (or more) compact laser power and 2-3 fold increase in pointing ability. With these developments, one can increase the distance from 2 AU for ASTROD to 10 AU (2×5 AU) and the spacecraft would be in orbits similar to Jupiter's. Four spacecraft would be ideal for a dedicated gravitational-wave mission (Super-ASTROD)." At that time, three spacecraft forming a triangle in the orbit plane of Jupiter together with one nearly perpendicular to this orbit plane ranging coherently (interferometrically) with one another were envisaged. This orbit configuration also has some capability of testing relativistic gravity and mapping the outer solar system. To strengthen the capability of testing relativistic gravity further, in the present version, we add a spacecraft near Earth-Sun L1/L2 point to tie the outer-solar-system ephemeris to the inner-solar-system ephemeris more accurately (This is also good for increasing the sensitivity of stochastic gravitational wave detection due to increased links.); the current ephemeris accuracy of outer solar system is about 10-1000 km while that of inner solar system is 1-1000 m, to improve the outer solar-system ephemeris with respect to inner solar system, this more precise link is needed. The baseline becomes 4+1 spacecraft. To make it flexible but still keep the ingredients, the mission concept has 3-5 large-orbit spacecraft (~5 AU) together with 1 Earth-Sun L1/L2 point spacecraft ranging coherently with one another. We also add a pulsed laser link between the off-Jupiter-orbit-plane spacecraft and ground laser stations for testing relativistic gravity, link to the inner solar system and optical communication, and for robustness of the mission.

### 2.1. Orbit

The baseline mission is to have four large-orbit spacecraft and one L1/L2 point spacecraft. The four large-orbit spacecraft will depart Earth with a velocity ~10 km/s in a direct-to-Jupiter orbit for Jupiter swing-by. The launch opportunity is every year. (If a launcher with less rocket power is chosen, ΔV-EGA orbit with an Earth-gravity-assistance (EGA) before reaching Jupiter can be considered.)

### 2.2. Payload and spacecraft

To achieve the objectives of the mission, the basic payload consists of
- (i) 15 W CW lasers,
- (ii) Pulsed laser and event timer,
- (iii) Optical clock, optical comb and frequency synthesizer,
- (iv) Telescope (40-50 cm $\varphi$) and optics,
- (v) Inertial sensor/accelerometer,
- (vi) Drag-free control and micro-Newton thrusters.

The spacecraft include
- (i) Radioisotope Thermoelectric Generators (RTGs),
- (ii) Two low-gain antennas for LEOP (Launch & early orbit phase):,
- (iii) Ka band communication,
- (iv) Propulsion module.

Since the solar radiation at 5 AU is 25 times less than at 1 AU and is not enough for power consumption, RTGs have to be used for power generation. Ka band communication is desired for its compact size. A propulsion module is needed for orbit manipulation. The processing and management of the spacecraft functions and payload systems are within the capabilities of the RAD 6000-SC (Radiation Hardened IBM RISC 6000 Single Chip) standard system; one (or an equivalent) system for the payload and one (or an equivalent) for spacecraft will be used.

## 3. Sensitivity to gravitational waves

A typical strain sensitivity curve for a fixed period of integration for a space interferometric GW detector consists of three frequency regions; the antenna response region, the flat region, and the acceleration noise region. Toward high frequency region, once the armlength $L$ exceeds half the GW wavelength ($\lambda_{GW}$), the antenna response rolls down roughly as $\lambda_{GW}/L$, i.e., as $f^{-1}$. Hence, the sensitivity curve rises and the curve shifts to the left proportional to $L$. In the middle flat region, the sensitivity curve is dominated by the white photon shot noise. In the lower frequency region, the curve is dominated by the acceleration noise and it rises with some inverse power of frequency. As armlength $L$ increases, the sensitivity curve is shifted downward since the corresponding strain noise is inversely proportional to $L$.

The intensity of a stochastic primordial background of GWs is usually characterized by the dimensionless quantity

$$\Omega_{GW}(f) = (1/\rho_c)\,(d\rho_{GW}/d\log f), \qquad (4)$$

with $\rho_{GW}$ the energy density of the stochastic GW background and $\rho_c$ the present value of the critical density for closing the universe in general relativity. The minimum detectable intensity of a stochastic GW background $\Omega_{GW}^{min}(f)$ is proportional to detector noise power spectral density $S_n(f)$ times frequency to the third power [4, 8]. That is

$$h_0^2\,\Omega_{GW}^{min}(f) \sim \text{const.} \times f^3\,S_n(f), \qquad (5)$$

where $h_0$ is the present Hubble constant $H_0$ divided by 100 km s$^{-1}$ Mpc$^{-1}$. Hence, with the same strain sensitivity, lower frequency detectors have an $f^{-3}$-advantage over the higher frequency detectors. This is the main reason why ASTROD and Super-ASTROD can probe deep into inflation GW region. In figure 1, compared to LISA, ASTROD has 27,000 times (30$^3$) better sensitivity due to this reason. For Super-ASTROD there is an additional 125 (5$^3$) times gain in sensitivity (the dotted line in between the ASTROD and Super-ASTROD sensitivity curves) compared to ASTROD due to longer armlength if the laser power and $S_n(f)$ stays the same. However, for baseline Super-ASTROD, the laser power is increased by 15 fold, and hence, the shot noise floor for strain sensitivity is lowered by a factor of 3.87. In terms of noise power spectral density, it is lowered by a factor of 15. Hence, the total gain in sensitivity due to $f^3\,S_n(f)$ is a factor of 1875. In the study of GW background from cosmological compact binaries, Farmer and Phinney [10] showed that this background $\Omega_{gw}^{cb}(f)$ rolls off in the 1-100 μHz frequency region from 10$^{-13}$ at 100 μHz to 10$^{-17}$ level at 1 μHz. Therefore, Super-ASTROD will be able to detect this background and there is still ample room for detecting the primordial/inflationary GWs with optimistic amplitudes. For ASTROD, although the sensitivity curve extends well below the optimistic inflation line, the background from binaries [10] in the ASTROD bandwidth is as large as the most optimistic inflation signals. Signal separation methods should be pursued. If relic primordial/inflationary GW has a different spectrum than that cosmological compact binaries, their signals can be separated to certain degree. With this separation, Super-ASTROD will be able to probe deeper into the stochastic waves generated during inflation with less optimistic parameters. More theoretical investigations in this respect are needed.

## 4. Testing cosmological and gravitational theories

Dark matter and dark energy are two eminent issues in astrophysics, gravitation and cosmology. There are two basic ways in which the mysteries of dark energy and dark matter might be solved. One is the direct measurement and characterization of the energy and matter concerned. The second is to modify the present theory of gravity. Inflation is successful in explaining the origin of cosmological perturbations and the degree of cosmological homogeneity and isotropy. However, the origin of inflation is a mystery. Various gravitational theories have been proposed for inflation, dark energy and dark matter. There are four ways these theories can be tested/constrained: laboratory precision experiments, solar-system experiments/observations, astrophysical observations and cosmological observations. For example, in the Dadhich-Maartens-Papadopoulos-Rezania solution [11] of the brane world models [12], the numerical values of the bulk tidal parameter and of the brane tension are constrained by the classical tests of general relativity in the solar system [13].

In a five-dimensional braneworld model developed by Dvali, Gabadadze and Porrati (DGP gravity) [14], the standard model (matter) interactions are constrained to a four-dimensional brane while gravity is modified at large distances by the arrested leakage of gravitons off our four-dimensional universe. DGP gravity has a crossover scale $r_c \approx 5$ Gpc, above which gravity becomes 5-dimensional. The model is able to produce cosmic acceleration without invoking dark energy. Lue and Starkman [15] showed that orbits near a mass source suffer a universal anomalous precession $d\omega/dt$ as large as ±5 μas/year, dependent only on the graviton's effective linewidth and the global geometry of the full, five-dimensional universe

$$|d\omega/dt| = 3c/8r_c = 5 \times 10^{-4} (5\text{Gpc}/r_c) \text{ arcsec/century}. \tag{6}$$

Iorio extended this equation to second order in eccentricity and used solar-system observations to constrain the anomalous gravitational effects [16]. Battat, Stubbs and Chandler [17] noticed that single point measurement uncertainties in the ranging data to Mercury and Mars are 10 m and 5-40 m, respectively, and for DGP-like precession the constraint is $|d\omega/dt| < 0.02$ arcsec/century. Therefore at the level of 0.02 arcsec/century, there is no evidence for a universal precession in excess of general relativity prediction.

One reason that the present constraints from the planetary motions are so relaxed is that they are nearly coplanar and for coplanar motion, universal precession can not be detected using relative motions. Super-ASTROD has one spacecraft orbit nearly vertical to the ecliptic plane and is ideal for this measurement. Two-wavelength laser ranging through the atmosphere of Earth achieved 1 mm accuracy [1, 18]. With a single point ranging accuracy of 1 mm using pulse ranging, the DGP effect of 180 m (for a mission of 10 years: $5 \times 10^{-5}$ arcsec × 4.8 × $10^{-6}$ rad/arcsec × 5 AU ≈ 180 m) for Super-ASTROD can be measured to $10^{-4}$ or better. For Super-ASTROD, 2nd order eccentricity effect in DGP theory can also be measured. This is an example of the capability of testing relativistic gravity.

## 5. Mapping the outer solar system

The Super-ASTROD spacecraft have larger orbit (~ 5 AU from the Sun), and thus are influenced more by the gravitating sources of outer solar system. During Jupiter flyby, the gravity field and multipole moments of Jupiter can be measured precisely. If during the transfer orbit, there is an Earth flyby, the Earth field can be measured and an additional link of Earth-Moon system frame to solar-system frame can be established. Moreover, the total mass of asteroids can be estimated from outside in addition to the estimation from the motion of the inner solar system. This may give a more precise and more consistent value. The perturbation from Kuiper-Belt Objects (KBOs) might also be measured to give a total mass estimation for KBOs.

From Table 1, anomalous Pioneer acceleration can be measured precisely by ASTROD I and ASTROD. This acceleration can also be measured precisely by Super-ASTROD from a larger distance more comparable to the distance at which Pioneer measured the effect.

## 6. Outlook

Super-ASTROD with Jupiter-size orbits ranging with each other has scientific objectives of exploring primordial and background gravitational-waves with frequency from 0.1 μHz to 1 mHz, testing relativistic theories of gravity and cosmological models, and mapping the solar-system mass distribution and dynamics. With current technology, the basic ingredients are there and the technology development will be mature for implementation of Super-ASTROD after LISA and ASTROD. For the important issue of detecting primordial GWs, missions on both sides of LISA detection bandwidth are worth further studies.

**Acknowledgements**
We would like to thank the National Natural Science Foundation of China (Grant Nos 10778710 and 10875171) and the Foundation of Minor Planets of Purple Mountain Observatory for support. We are also grateful to the referees for critical and helpful comments on improving the manuscript.